\documentclass[]{spie}  

\usepackage[utf8]{inputenc}
\usepackage{graphicx}
\usepackage{ragged2e}
\usepackage{wrapfig} 
\usepackage{epstopdf}
\DeclareGraphicsExtensions{.pdf,.png,.jpg}
\usepackage{multirow}
\usepackage{varioref}
\usepackage{multicol}
\usepackage{float}
\usepackage{url}
\usepackage{rotate}
\usepackage{gensymb}
\usepackage{subcaption}
\usepackage{amsmath,amsfonts,amssymb}
\usepackage{graphicx}
\usepackage[colorlinks=true, allcolors=blue]{hyperref}

\title{NPF: mirror development in Chile}

\author[a,b]{Sebastián Zúñiga-Fernández}
\author[a,b]{Amelia Bayo}
\author[a,b]{Johan Olofsson}
\author[b,c]{Leslie Pedrero}
\author[b,c]{Claudio Lobos}
\author[c]{Elias Rozas}
\author[b,c]{Nicolás Soto}
\author[a,b]{Matthias Schreiber}
\author[d,c,b,f]{Pedro Escárate}
\author[c,b]{Christian Romero}
\author[c,b]{Hayk Hakobyan}
\author[b,e]{Jorge Cuadra}
\author[b,c]{Cristopher Rozas}
\author[g]{John D. Monnier}
\author[f]{Stefan Kraus}
\author[h]{Mike J. Ireland}
\author[b]{Pedro Mardones}

\affil[a]{Instituto de Física y Astronomía, Universidad de Valparaíso, Avenida Gran Bretaña 1111, Casilla 5030, Valparaíso, Chile}
\affil[b]{Núcleo Milenio de Formación Planetaria (NPF), Valparaíso, Chile}
\affil[c]{Centro Científico Tecnológico de Valparaíso (CCTVal), Universidad Técnica Federico Santa María, Av. España 1680, Valparaíso, Chile}
\affil[d]{Large Binocular Telescope Observatory (LBTO), Safford, Arizona 85546, USA}
\affil[e]{Pontificia Universidad Católica de Chile, Av Libertador Bernardo O'Higgins 340, Santiago, Región Metropolitana, Chile}
\affil[f]{School of Physics, University of Exeter, Exeter, EX4 4QL, UK}
\affil[g]{Department of Astronomy, University of Michigan, Ann Arbor, Michigan 48109, USA}
\affil[h]{Research School of Astronomy and Astrophysics, Australian National University, Canberra, ACT 2611, Australia}
\affil[f]{Instituto de Electricidad y Electr\'onica, Universidad Austral, Valdivia, Chile}

\authorinfo{Further author information: (Send correspondence to A.B.)\\  S.Z.: E-mail: sebastian.zuniga@postgrado.uv.cl\\A.B.: E-mail: abayo@npf.cl, Telephone: 56 32 250 8311}

\begin{document} 
\maketitle

\begin{abstract}
In the era of high-angular resolution astronomical instrumentation, where long and very long baseline interferometers (constituted by many, $\sim$20 or more, telescopes) are expected to work not only in the millimeter and submillimeter domain, but also at near and mid infrared wavelengths (experiments such as the Planet Formation Imager, PFI, see Monnier et al. 2018 for an update on its design); any promising strategy to alleviate the costs of the individual telescopes involved needs to be explored.

In a recent collaboration between engineers, experimental physicists and astronomers in Valparaiso, Chile, we are gaining expertise in the production of light carbon fiber polymer reinforced mirrors. The working principle consists in replicating a glass, or other substrate, mandrel surface with the mirrored adequate curvature, surface characteristics and general shape. Once the carbon fiber base has hardened, previous studies have shown that it can be coated (aluminum) using standard coating processes/techniques designed for glass-based mirrors. 

The resulting surface quality is highly dependent on the temperature and humidity control among other variables. Current efforts are focused on improving the smoothness of the resulting surfaces to meet near/mid infrared specifications, overcoming, among others, possible deteriorations derived from the replication process. In a second step, at the validation and quality control stage, the mirrors are characterized using simple/traditional tools like spherometers (down to micron precision), but also an optical bench with a Shack-Hartman wavefront sensor.  

This research line is developed in parallel with a more classical glass-based approach, and in both cases we are prototyping at the small scale of few tens of cms. We here present our progress on these two approaches.
\end{abstract}


\section{INTRODUCTION}
\label{sec:intro}  

Since about a decade ago, Chilean Astronomy is experiencing a golden era of tremendous growth and prosperity. Research centers performing world-class investigations have formed at several institutions and Chile has become an important country in terms of astronomical research. This development during the last decades has become possible thanks to the amazing facilities installed in the North of Chile by international organizations and the right of Chilean institutions to use these facilities. We believe the time has come to convert Chile from a nearly pure ``user" of these high-end astronomical equipment to a country that actively participates in the construction and development of new astronomical facilities. We believe that this next step is ideally based on research excellence in one particular area of astronomy and on the involvement of Chilean researchers in a large international project.

\subsection{The next challenge}

Planet formation is indisputably one of the most exciting topics of modern astronomy. Since the discovery of the first planet around a sun-like star in 1995 \cite{Mayor95}, planet hunting has become a main discipline of astronomical research. Thanks to missions like Kepler and CoRoT, and much longer astrometric and radial velocity campaigns, we now have thousands of fully-formed confirmed planets with a wide range of orbital and physical characteristics. We have learned that nature is able to produce a huge variety of planetary systems which underlines the complexity of the formation mechanisms; making the search for planets ``in the forming", a total must to progress in this field. 

State-of-the-art observing facilities such as SPHERE (working in the optical/near infrared) and ALMA (working at radio wavelengths) are opening new avenues in this respect and new discoveries in the field of planet formation are published almost on a weekly time scale. However, neither instrument/facility, by design, offer the possibility of spatially resolving the Hill sphere of forming planets (their gravitational radius of action).   

This science case sets the basis for the PFI project (www.planetformationimager.org). We refer the reader to Monnier at al. 2018 for an update on the status of the project, but in short, PFI is an ambitious initiative from the global scientific community designed to spatially resolve the Hill sphere of forming planets at a typical distance of 140 pc. For reference, the Hill sphere of Jupiter has a size of 0.35 au. 
Simulations have been carried out by the PFI team to estimate the optimal wavelength range where the proto-planet emission should dominate over contamination from surrounding material at any stage of the formation process ($\sim$0.1-100 Myrs). These simulations show that the thermal/mid-infrared (MIR) regime (in particular $\sim$3-10 micron) is the ``sweet spot”. The resulting combination of wavelength range and angular resolution requirements ($\sim$0.2 mas at 10 microns), yields $\sim$km baselines for a thermal/MIR interferometer that will consist of a dozen of individual telescopes each one with diameters between 2 and 4\,m.

Mirrors represent a significant fraction of the total costs of a telescope. From the smallest scale, the cost to buy and import a 70cm mirror corresponds to $\sim$22\% of the total cost of such size telescope. In addition, mirror prices increase significantly with size. Local development of the technology and know-how to build high-quality 1m segments (to be assembled in 2-4m structures) could alleviate significantly the final cost for any telescope to be installed in Chile. Therefore, cost-effective design and manufacturing of relatively small mirrors or segments, presents a fascinating technological challenge,
which is the foundation for our ``N\'ucleo Milenio de Formaci\'on Planetaria'' (NPF\footnote{\url{http://www.npf.cl/}}) collaboration, in Valpara\'iso, Chile. 
Valpara\'iso offers ideal conditions since the alliance of Universidad de Valpara\'iso (UV) and Universidad T\'ecnica Federico Santa Mar\'ia (UTFSM) within NPF provides on the one hand a strong group studying planet formation, and, on the other hand, a group already heavily involved in mirror design and manufacturing for rougher precision and smaller scale projects. 

In this manuscript we present the two approaches (carbon fiber technology and glass-based solution) that we are following, and summarize the preliminary results of our prototyping.
\section{Carbon fiber-based mirrors}
\label{sec:CF_mirror}

\begin{figure}
	\centering
	\includegraphics[width=0.7\textwidth]{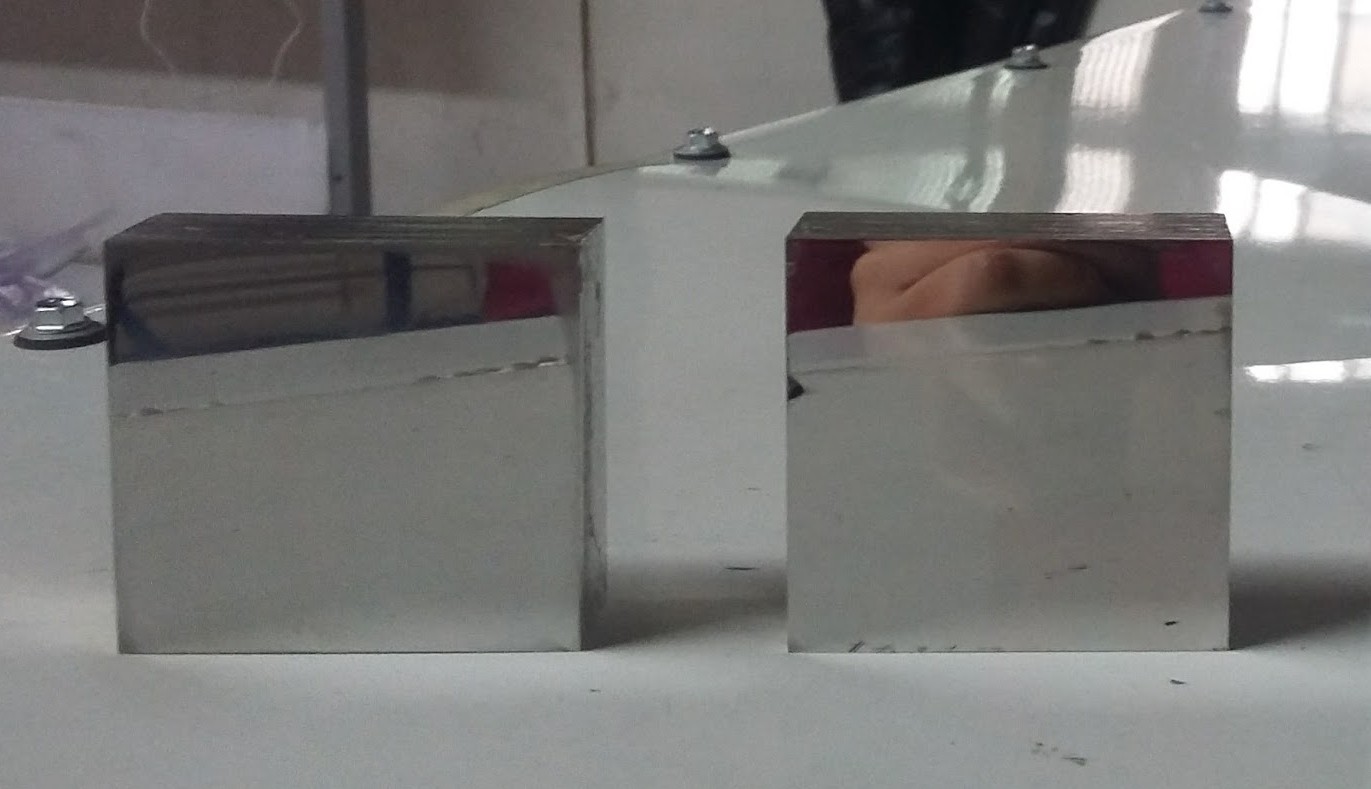}
	\caption{Polished steel mandrels used to create flat CF based prototype mirrors.}
	\label{fig:mandrel_proto}
\end{figure}

\begin{figure}
	\centering
	\includegraphics[width=0.6\textwidth]{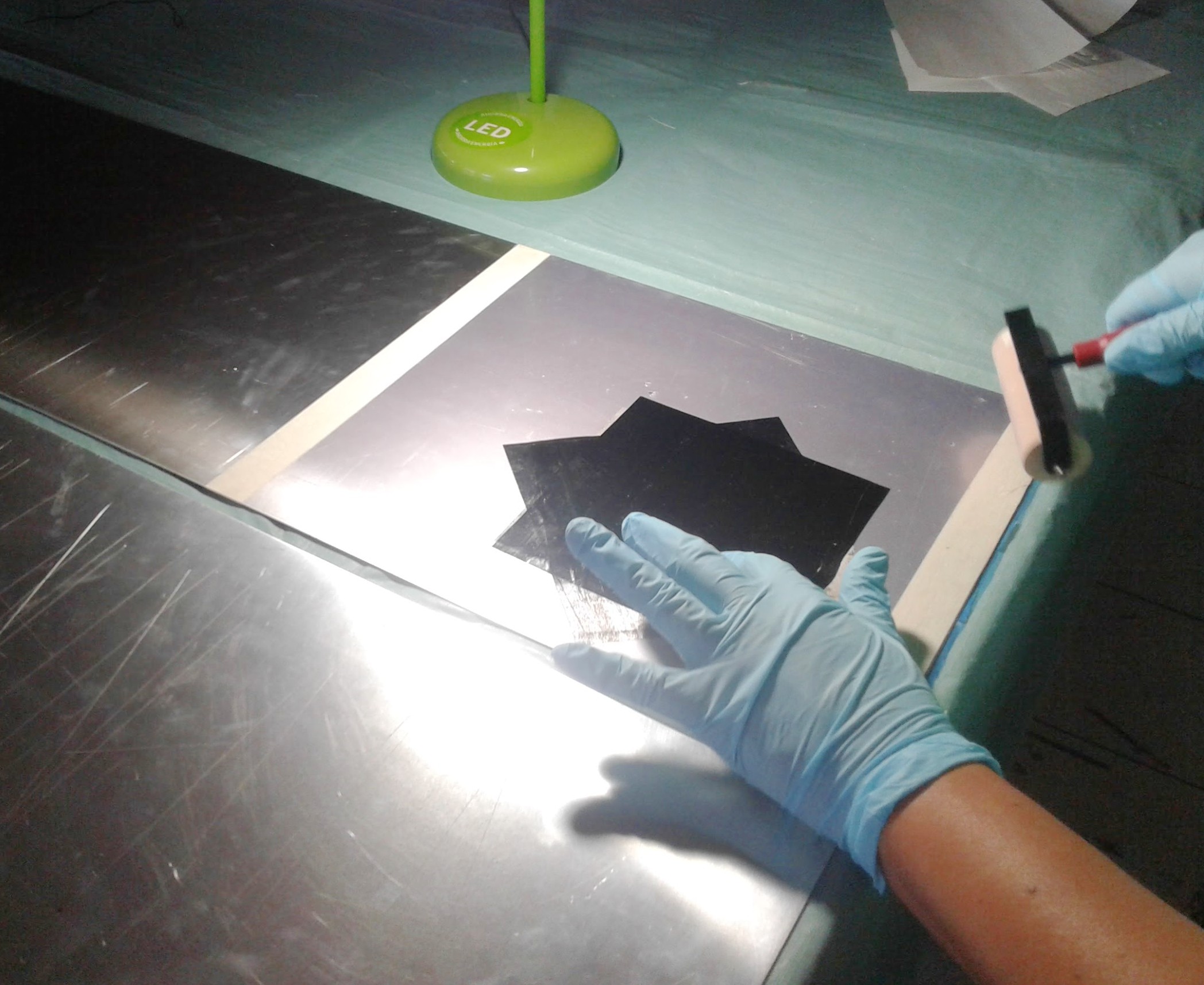}
	\caption{Lay up process for small prototypes. Every CF layer is placed with a 45$^{\circ}$ angle rotation with respect to the previous one.}
	\label{fig:layup}
\end{figure}

The main step to shape the carbon fiber (CF, hereafter) bases relies on the well known technique of vacuum bagging. This technique is used for modeling small parts for cars, boats, and even space mission applications. In the production of a CF mirror, one key component is the mold/mandrel used. This mold can be made of different materials, such as stainless steel, marble, ceramic materials, and even astronomy-class glass (see Fig.\,\ref{fig:mandrel_proto}). The fiber sheets are cast onto this mold layer by layer (four layers in our case), alternating the orientation of the fibers, to improve the overall smoothness and mechanical stability. The lay up process needs a rigorous manipulation to avoid air, humidity or wrinkles between each layer of CF over the mandrel. The process is, therefore, extremely sensitive to environmental conditions (see Fig. \ref{fig:layup}).

\begin{figure}[H]
	\centering
	\includegraphics[width=0.7\textwidth]{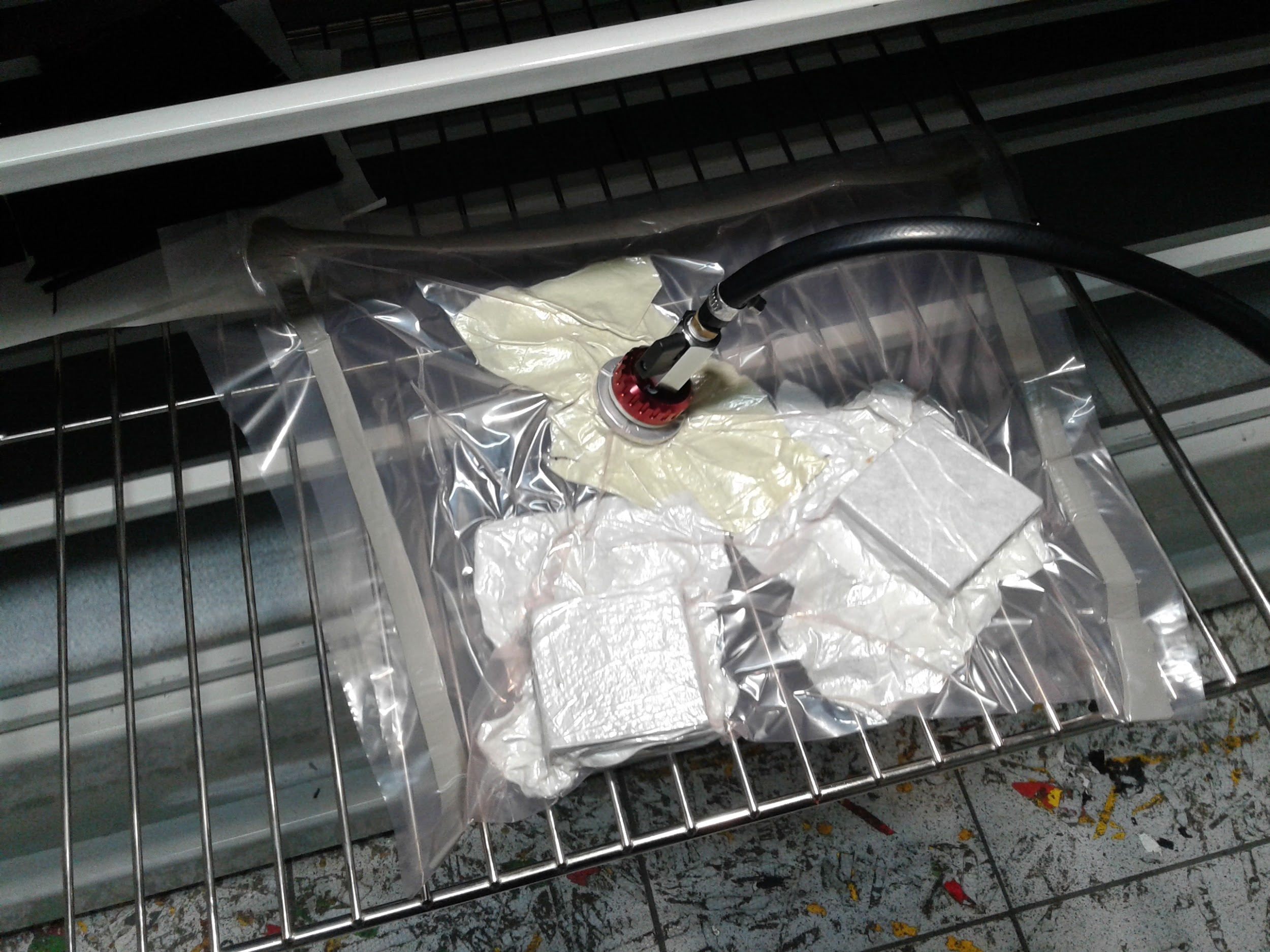}
	\caption{Vacuum bagging technique applied to small prototypes. The pressure is maintained during the whole curing process inside the oven.}
	\label{fig:vacuum_bagging}
\end{figure}

Once the lay up process is completed, the CF is cured at a temperature ranging between 120 and 140$^{\circ}$ C. During the process, external forces are applied so that the CF is ``pressed” against the mold. A possible way to achieve this is to introduce a vacuum bag inside the curing oven (see Fig. \ref{fig:vacuum_bagging}), which in general will only provide one atmospheric pressure, and/or with higher pressure-inflicting devices such as autoclave ovens.  

The next step consists of cooling to room temperature the substrate and then proceed with the lifting-off process. The main hypothesis to be tested at this stage is the accuracy of the ``replicated/mirrored" surface. One must note that in every step of the manufacturing process a set of variables, that influence the surface quality, is added. For example, the handler's expertise, and environmental control are key aspects. However, in our experience, even with great care and clean environment, the relatively low pressure of $1$\,bar being currently applied in the curing process is the main limiting factor. Our replicated pieces present deterioration in the surface quality of the order of a factor $\sim$10.

The next step in this line of prototyping is therefore to acquire an autoclave oven to work at the $\sim$3 bar regime.

\section{Glass-based mirrors}
\label{sec:glass_mirror}

In the case of glass-based mirror manufacturing, a first vital re-shaping (outer shaping) step provides the base, a so-called blank. It is followed by an iterative process of successive grinding, polishing and measuring until the surface fulfills the requirements for its (spherical) shape and smoothness. 
In addition to the aforementioned three stages, a final ``figuring" step needs to be carried out to modify the surface curvature from spherical to, for example parabolic or hyperbolic. In the rough grinding step the main goal is to ``carve out" the expected curvature of the mirror, that will obviously depend on the overall focal ratio. 

In general, the key idea behind all these steps is that one can remove material from the surface as long as one uses another material with higher hardness. For these steps telescope manufacturers often use silicon carbide powders (a.k.a. carborundum). To be effective, the abrasive needs to be rubbed against the glass with a significant amount of pressure which is applied using a customized tool. As a rule of thumb, this tool should be of about the same size and weight as the blank. 

\begin{figure}
\begin{subfigure}{.5\textwidth}
  \centering
  \includegraphics[width=.8\linewidth]{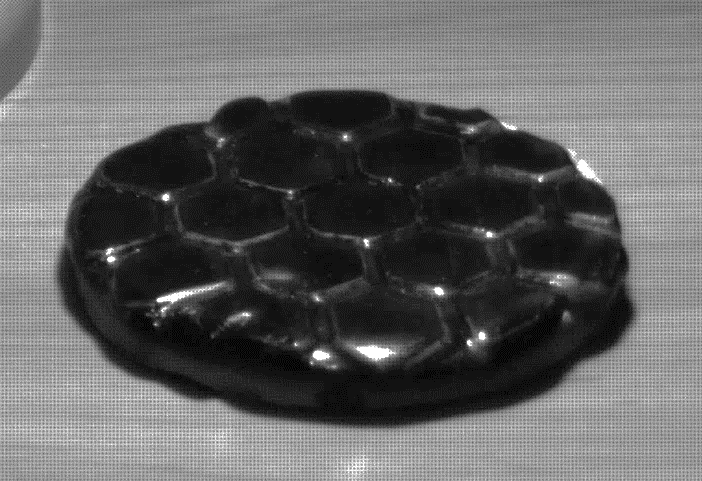}
  \label{fig:sfig1}
\end{subfigure}%
\begin{subfigure}{.5\textwidth}
  \centering
  \includegraphics[width=.87\linewidth]{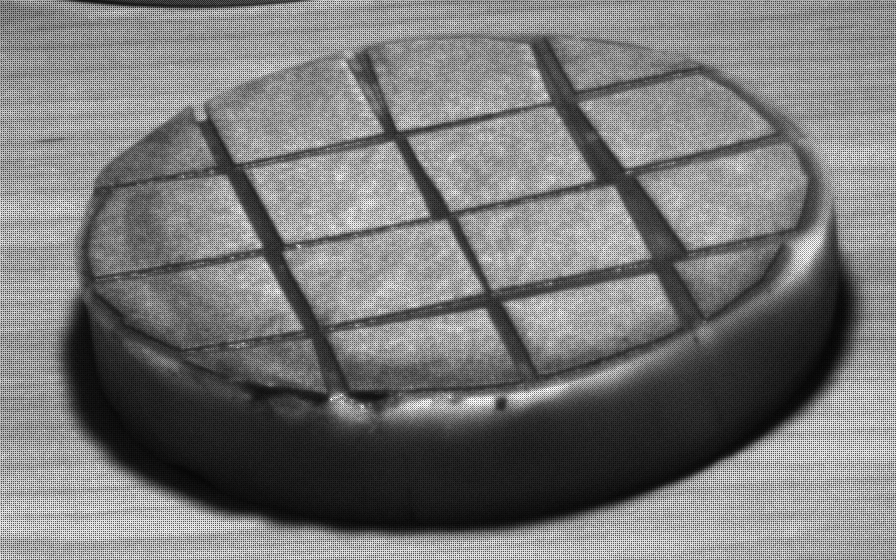}
  \label{fig:sfig2}
\end{subfigure}
\caption{Examples of grinding tools made in our laboratory. An optical pitch tool on the left and a marble based tool on the right.}
\label{fig:grinding_tool}
\end{figure}

In our particular case, we built a semi-automatic machine for rough grinding. This simple machine is based on the eccentric movement of an oscillating arm combined with the rotation of the blank and the grinding tool (Fig. \ref{fig:grinding_tool}). 

Secondly, the fine grounding serves one main purpose: to remove the scratches and holes left by the rough grinding while maintaining the overall shape of the surface. In order to achieve this goal, a finer abrasive is used, continuing to finer and finer grain sizes until all the holes and scratches have been removed.

The goal of the final polishing process, is to remove the rough surface left during the fine grinding, to therefore obtain a smooth reflective surface. During this step instead of creating fractures in the glass, the grinding agent (like cerium oxide or diamond) will act as shears removing peaks on the surface (at microscopic scales).

Finally, as previously stated, the figuring step consists of measuring and re-shaping the perfectly spherical surface into the requirements of the specific mirror (e.g. parabolic or hyperbolic). This task can, in our experience, be one of the most time demanding steps, where the expertise of the person handling the process is crucial.

\section{results}
\label{sec:results}
During the first months of our project, which started in September 2017, we have focused on gaining experience and training new people on different steps of the manufacturing processes, including testing new materials, tools, quality control methods and the effects of environmental conditions.
The results of these tests, the description of the produced equipment, and the justification for upgrades and new equipment purchases are detailed in this section.

One key point in which we have invested time and resources has to do with developing and acquiring instrumentation for polishing, measuring and coating. 

Our current polishing machine, operational for small prototypes, was built in-house, however,  we recently bought a 1 meter, semi-automatic, state-of-the art polishing machine. This polishing machine will not only allow us to produce high quality glass molds (resulting by our experiments in much better surfaces to be replicated than those of other material), but also to experiment with finishing polishing techniques of the carbon fiber replicas with metallic (nickel, for example) powder layers.

\begin{figure}[H]
	\centering
	\includegraphics[width=0.8\textwidth]{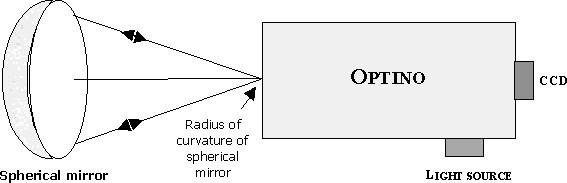}
	\caption{Shack-Hartmann based wave front sensor diagram for quality control measurement. It allows surface reconstruction with $\sim 10$ nm resolution at $\lambda=635$ nm. With some minor modifications this configuration can be used for parabolic and flat mirror tests (photo credit from Spot-Optics).}
	\label{fig:optino}
\end{figure}

\begin{figure}[H]
\begin{subfigure}{.5\textwidth}
  \centering
  \includegraphics[width=.65\linewidth]{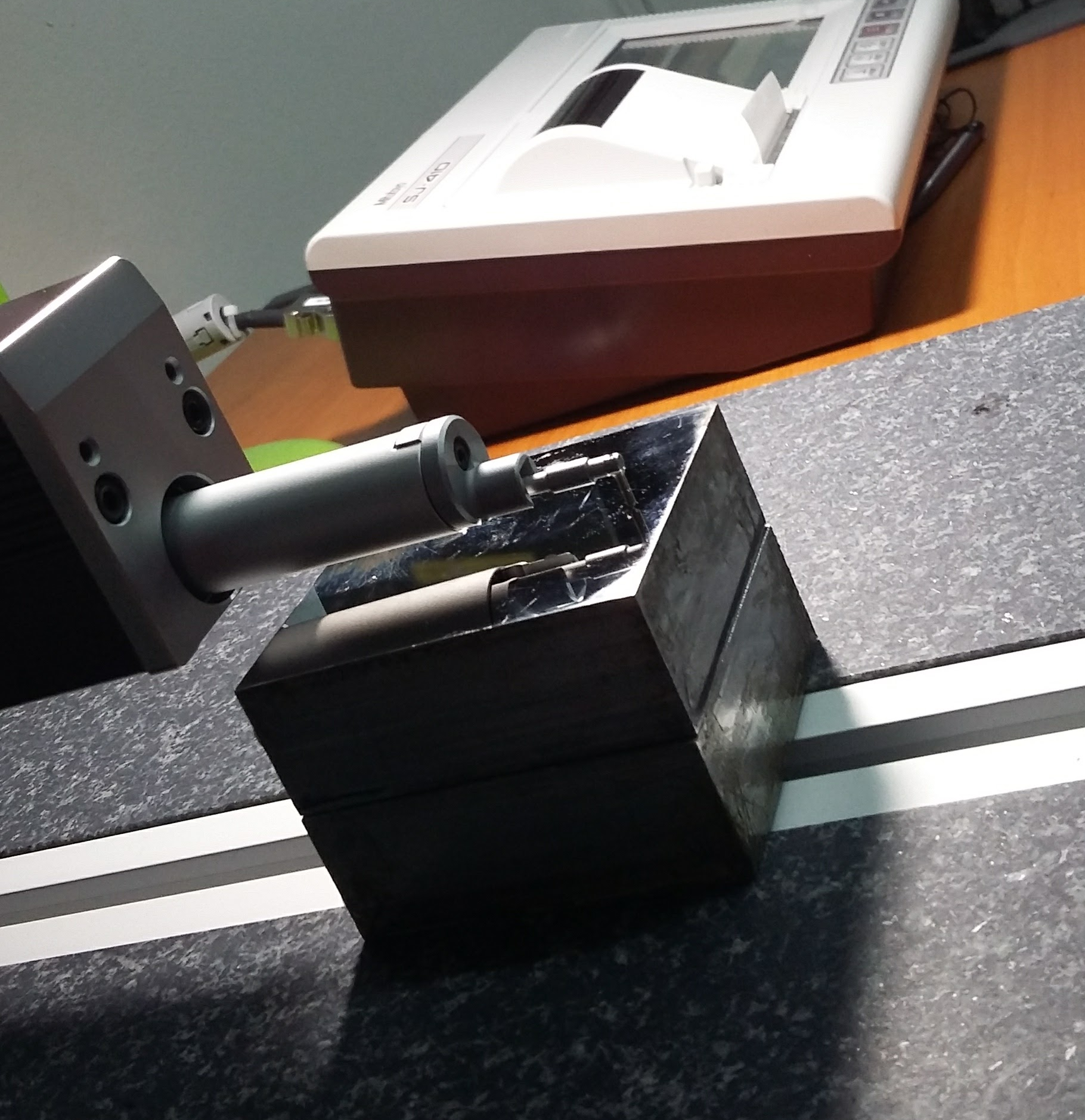}
  \label{fig:rugo}
\end{subfigure}%
\begin{subfigure}{.5\textwidth}
  \centering
  \includegraphics[width=.91\linewidth]{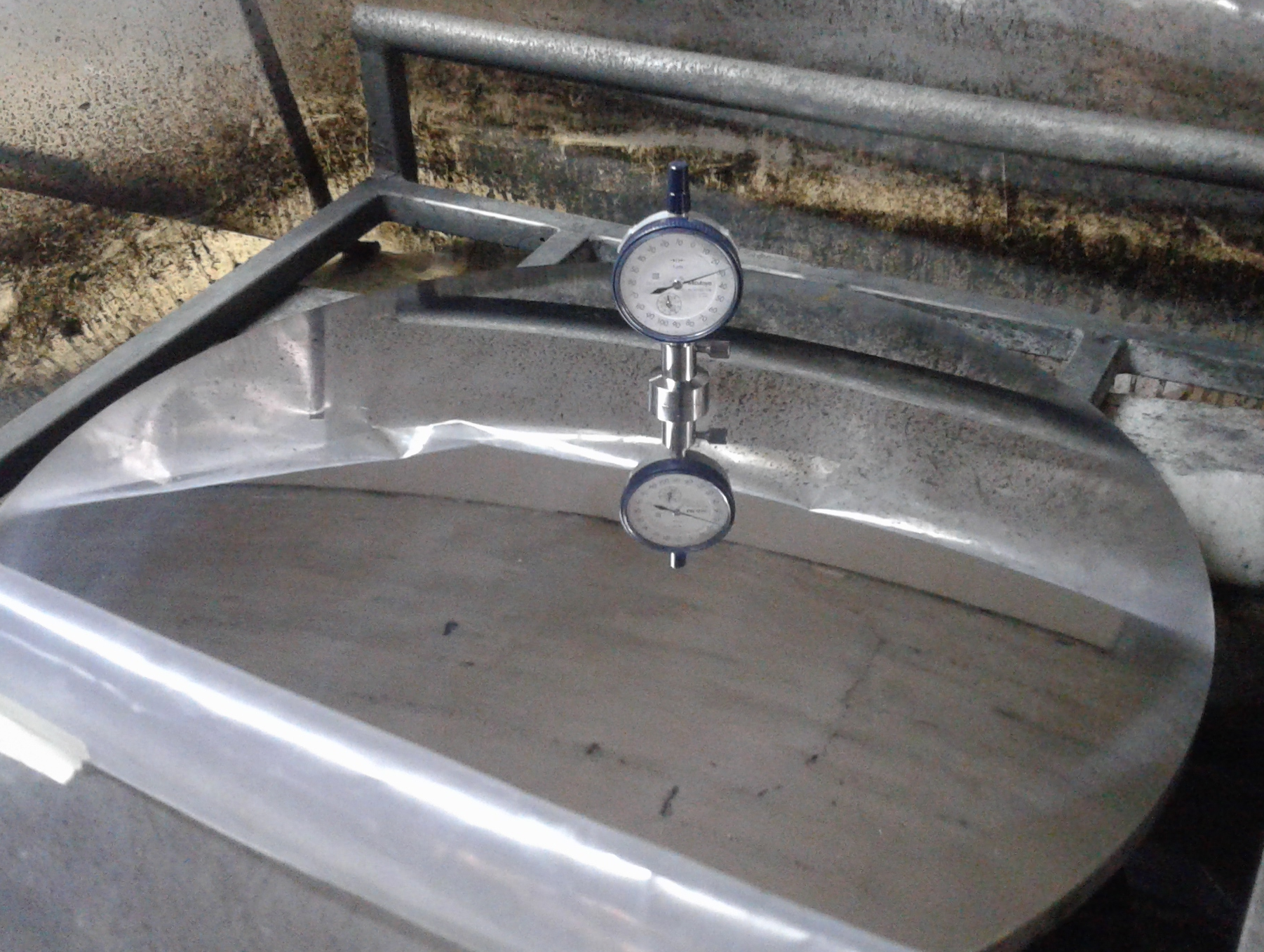}
  \label{fig:esfero}
\end{subfigure}
\caption{Contact measurement tools. Left panel: Surface roughness tester being applied over a small flat mandrel. Right: spherometer measuring over a 50 cm spherical mandrel. These instruments can measure down to precisions of $10-100$\,nm and 1 $\mu$m respectively.}
\label{fig:measure_tool}
\end{figure}

In addition, today at the laboratory we have equipment to characterize the mirror surfaces both optically (Fig. \ref{fig:optino}) and by contact (Fig. \ref{fig:measure_tool}). 

Finally, we also have a large two-channel sputtering chamber where it is possible to apply aluminum to the mirror surface (and simultaneously a protective layer). This large piece of equipment can hold mirrors of sizes up to 1.9 meters in diameter. 

\subsection{Carbon Fiber Based Mirrors}
\label{sec:result_CFmirror}
As mentioned before, every step in the CF mirror making process has been studied and tested consistently to improve the resulting mirror surface quality. The most notable tests and results for every step in the CF mirror manufacture are reported below.

\subsubsection{Mandrel polishing}

As described in Section\,\ref{sec:CF_mirror}, we are able to shape the CF mirror using the mandrel as a mold with the vacuum bagging technique. Naturally, if the shaping process is done properly, the better the mandrel is, the better the resulting mirror will be.
The carbon fiber technicians were trained in metal polishing techniques achieving significant improvements in the final surface quality of our steel-based mandrel. Despite the progress some issues still persist, mostly related to the spherical 50\,cm steel mandrel. On-going studies reveal that we need additional tools to achieve a precise curvature radius. Nevertheless the latest measurements indicate a roughness of the order of $20$\,nm. At this stage of the project, this is sufficient for us to continue developing flat and spherical mirrors, and to test also the next manufacturing steps at the prototype level. 

\subsubsection{Layup process and CF manipulation}

We are using linear prepreg (pre-impregnated) 
carbon fiber, and its manipulation must be done with extreme care. As examples of possible deterioration issues due to poor handling, the fibers can start separating, the carbon fiber can stick to itself, can bend, or even water can condense on it in case of prolonged manipulation in a humid environment. During the past year, the CF technicians have gained a lot of experience in properly manipulating the carbon fiber, and we purchased additional tools that facilitate the handling of this ``delicate" material. At the moment, we are focusing on upgrading the facilities to avoid damaging the carbon fiber when it is moved from the storage unit (which is cooled at $\sim -20 ^{\circ}$\,C) to the workspace. Improvements on the laboratory space are already on-going, including the implementation of a clean room with temperature (and other environmental variable) control.

\subsubsection{Vacuum bagging and release agent}

In the first stages of the project, the sealing of the vacuum bag was imperfect, which led to some leaking. This problem has been however solved now, conducting a more streamline process to maintain the adequate pressure in the bag. Obtaining release agents was also a logistical challenge, due to import restrictions, since these agents are commonly considered dangerous substances. Nowadays we are working with two release agents; one based on nano-particles (organic and ceramic version) from a Chilean company and the ``Frekote Nc'' from Locktite, as recommended by the \textit{``Associcao Latino-Americana de Materiais Compositos''}. The two release agents have shown promising results, but we will continue testing them to evaluate in more detail their respective performance.

\subsubsection{Curing oven and temperature control}

The curing process is carried out in a temperature controlled oven with size and temperature range best suited for our applications. It was nonetheless necessary to improve the heating curves to match the specific curing curves required for carbon fiber. This process was carried out using several thermocouples to obtain a proper characterization of the oven behavior over the course of a month. With those results, we updated the oven's settings to suit our needs, which greatly helped improving the final result of our prototypes.

\subsubsection{Structural support}

\begin{figure}[H]
	\centering
	\includegraphics[width=0.8\textwidth]{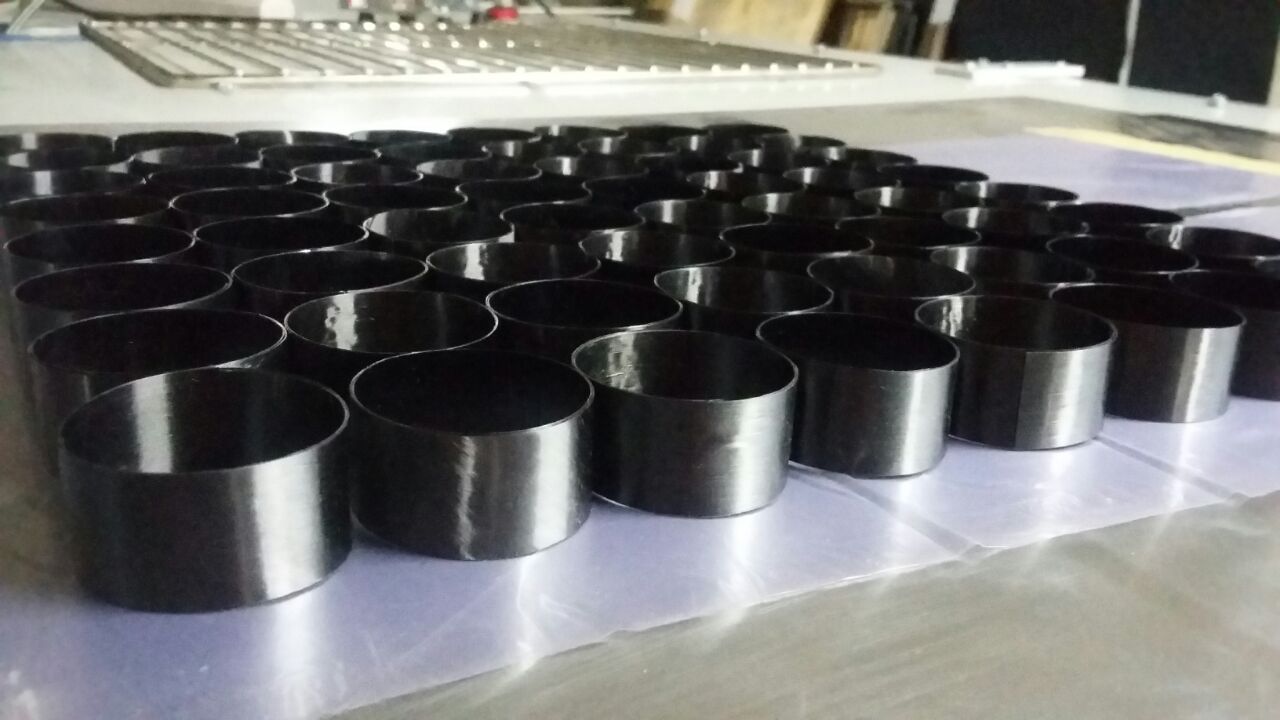}
	\caption{Carbon fiber honeycomb manufactured in our laboratory as a structural support for the mirror. This option is currently under investigation and further testing will soon be done using new molding tools.}
	\label{fig:CF_honey}
\end{figure}

\begin{figure}[H]
\begin{subfigure}{.5\textwidth}
  \centering
  \includegraphics[width=.9\linewidth]{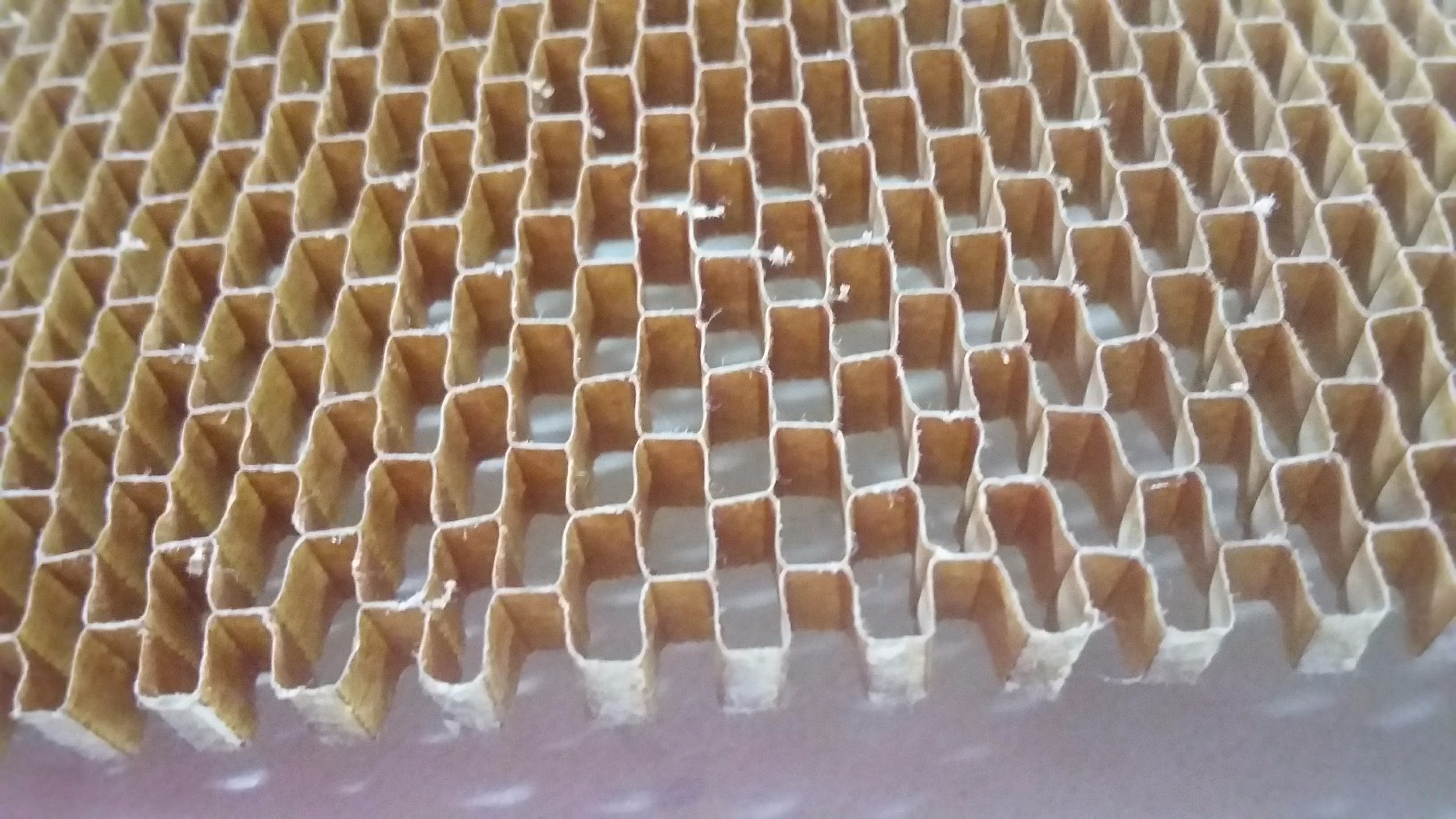}
  \label{fig:aramid}
\end{subfigure}%
\begin{subfigure}{.5\textwidth}
  \centering
  \includegraphics[width=.9\linewidth]{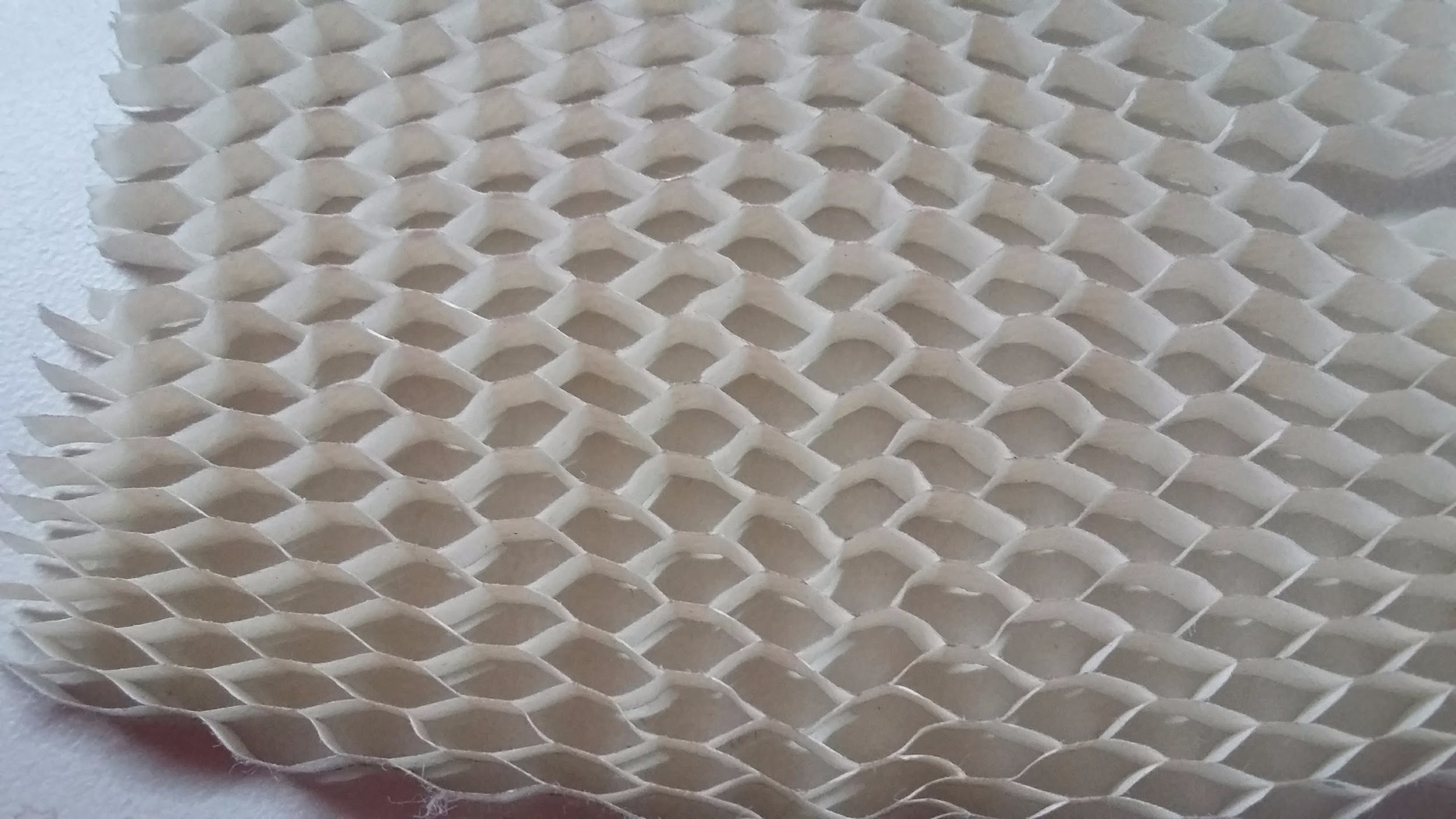}
  \label{fig:papel}
\end{subfigure}
\caption{The aramid honeycomb (left) and the paper honeycomb (right) are the structural support of the carbon fiber mirror.}
\label{fig:honeycomb}
\end{figure}

\begin{figure}[H]
	\centering
	\includegraphics[width=0.7\textwidth]{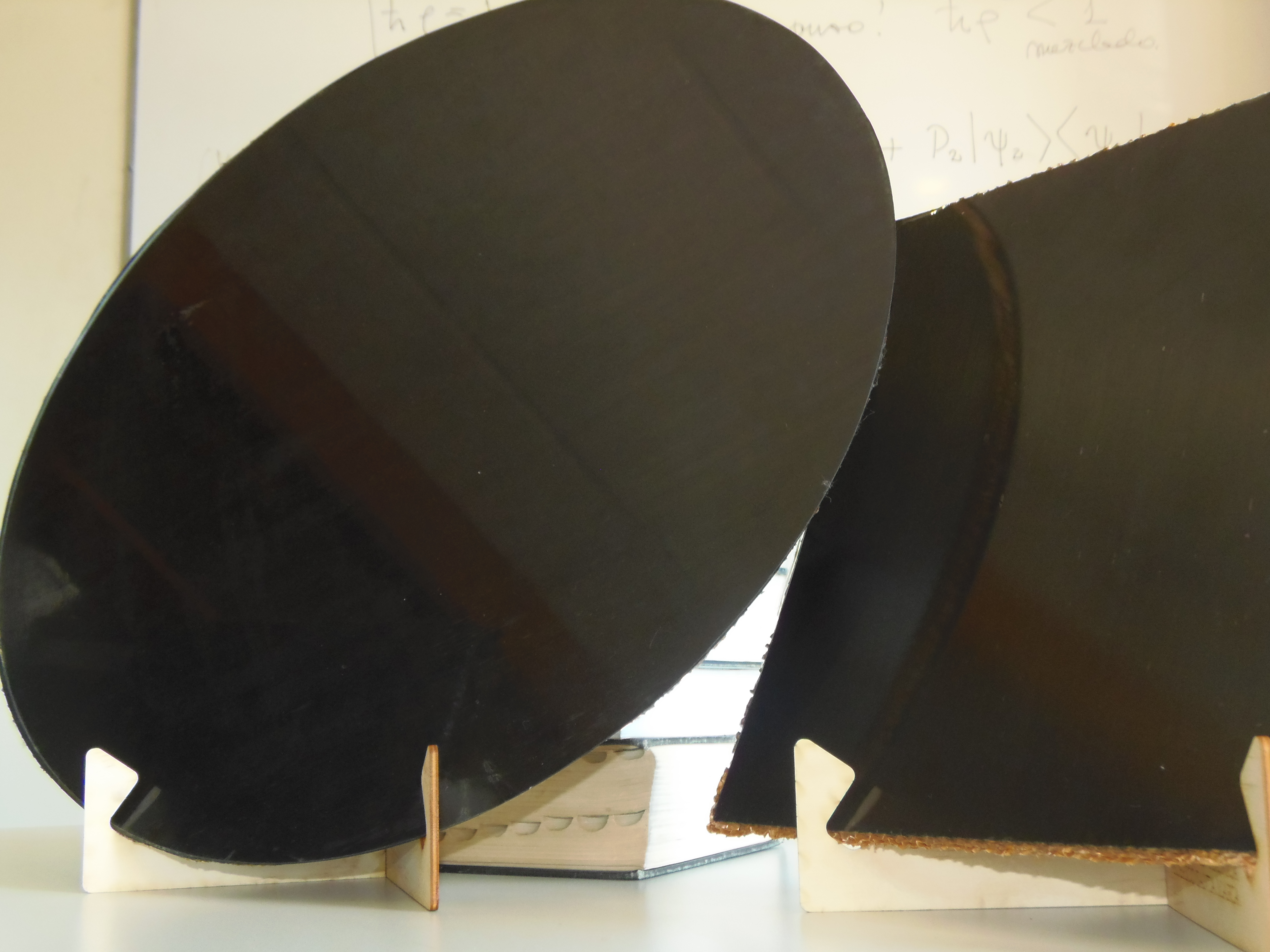}
	\caption{Spherical curvature CF prototypes with aramid honeycomb core using the 50 cm stainless steel mandrel.}
	\label{fig:CF_prototype}
   \end{figure}

Ideally, the carbon fiber lay up has to be performed on a stable, handleable, and lightweight structure. We are currently performing tests on different structures to keep the CF mirror in place. Initially, we were using a solid foam of polyurethane, but this structure led to some noticeable deformation in the final surface, as well as resin leakage, so it was rapidly discarded. Inspired by the literature, 
we developed a CF honeycomb structure in our laboratory (see Fig.\,\ref{fig:CF_honey}). We are still investigating this solution, and at the moment, we still need dedicated tools to improve the honeycomb manufacturing. As a result, we are not yet using this solution, but we are using either aramid or paper honeycombs instead (see Fig. \ref{fig:honeycomb}). These honeycombs originally presented (transferred) print-through  problems
, but after improving the curing process those problems were solved. This solution offers a good, lightweight, structural stability to the CF mirror (see ~\ref{fig:CF_prototype}).  

 \begin{figure}[H]
\begin{subfigure}{.5\textwidth}
  \centering
  \includegraphics[width=.95\linewidth]{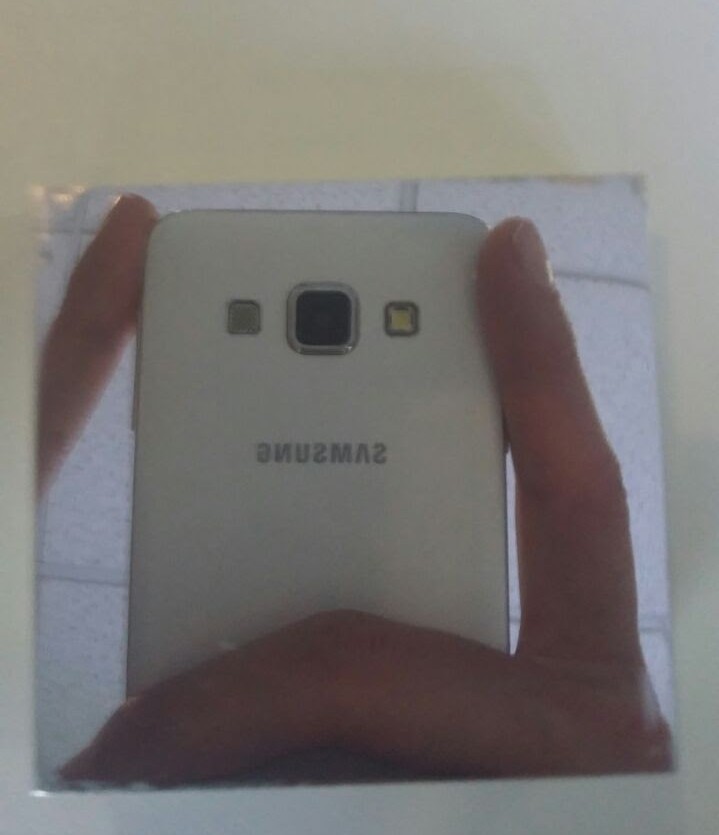}
\end{subfigure}%
\begin{subfigure}{.5\textwidth}
  \centering
  \includegraphics[width=.7\linewidth]{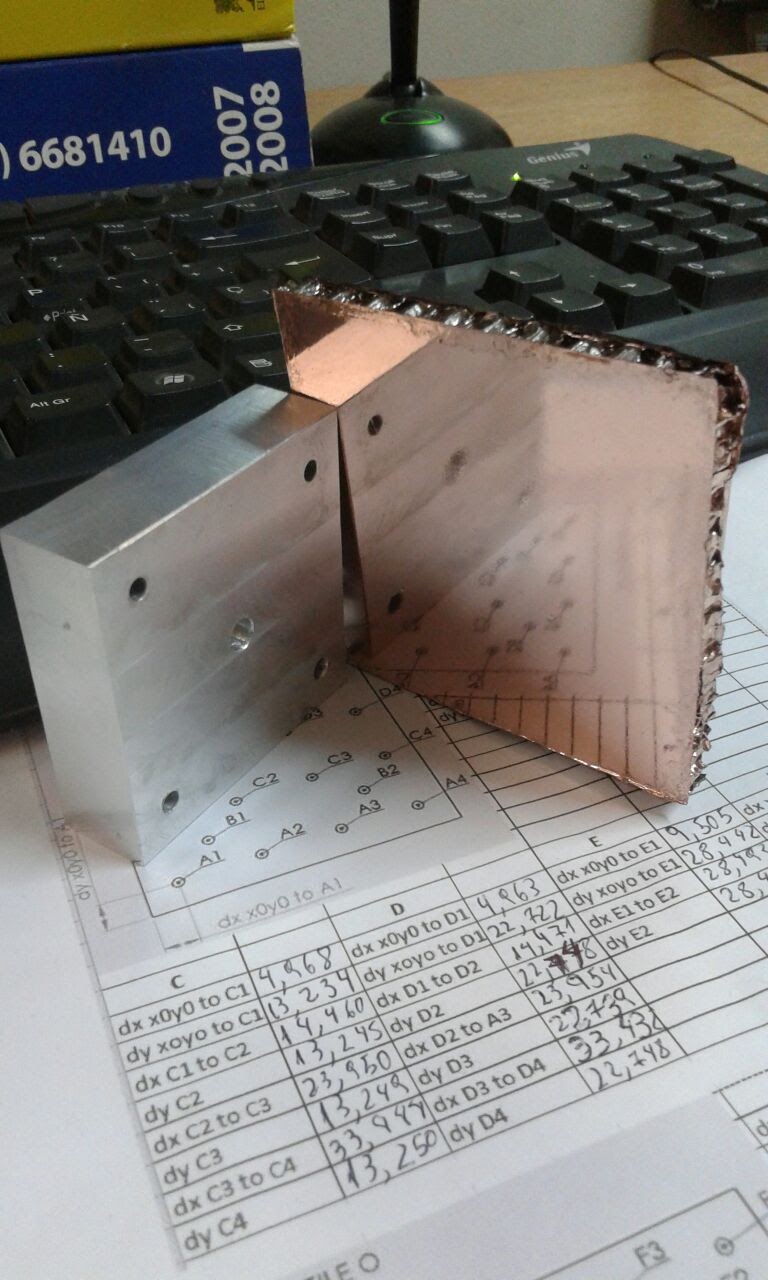}
\end{subfigure}
\caption{CF mirrors prototypes (10 cm, flat) with aluminum (left) and copper (right) deposited on their surfaces with a provisional Denton Vacuum sputter machine 
.}
\label{fig:sputter_results}
\end{figure}

\subsection{Glass Based Mirrors}
\label{sec:result_glassmirror}

On the long term, the motivation is to develop a complete polishing work-flow for glass mirrors of 1 meter, reaching astronomical quality (at worst tens of nm precision). Nowadays we are working on a 40 cm diameter mirror using different materials, techniques and abrasives in order to compare the results, gain expertise, and properly document the whole process.

\subsubsection{Polishing machine}

We carried out intensive research to identify the most suitable polishing machine. The main requirements for the machine were to be semi-automatic, to be ale to handle a one meter blank, and produce a spherical curvature and concave or convex surface. Our goal is to obtain a surface quality with an RMS roughness less than $20$\,nm and a wavefront smaller than $\lambda/10$, at a referential wavelength of $550$\,nm (i.e. above, in principle than the science PFI requirements). We selected the OPTO TECH HM1200 machine, considering the required technical specifications and the working experience of our expert on telescope manufacturing. Unfortunately we cannot show results with this machine yet, as it will not be part of the laboratory until the clean room is ready to host it.

   
\subsubsection{Mirror and tool materials}

\begin{figure}[H]
	\centering
	\includegraphics[width=0.6\textwidth]{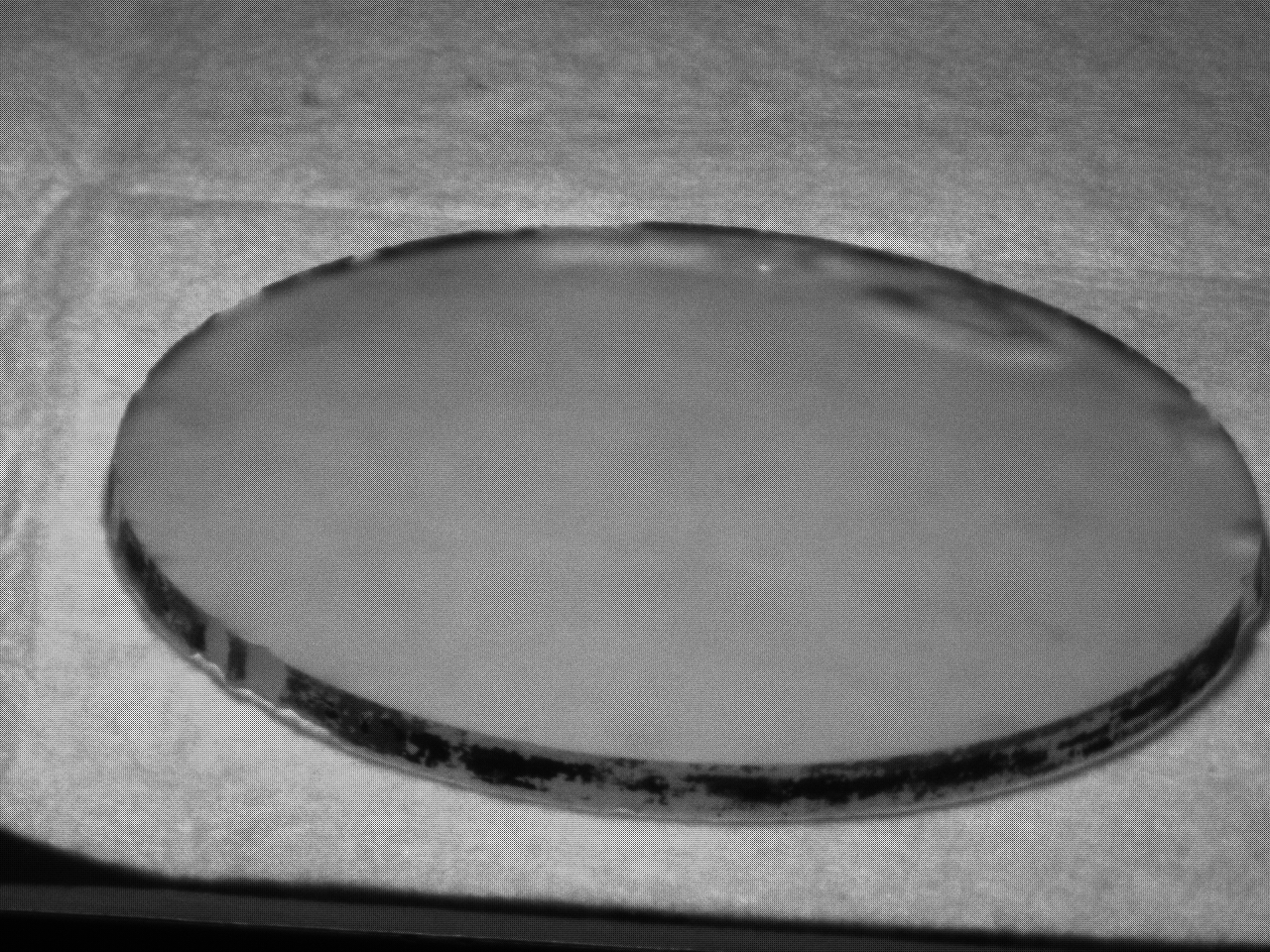}
	\caption{Silica 15 cm mandrel used for abrasive and polishing tests.}
	\label{fig:silica_mirror}
   \end{figure}

The material for the base of our glass-like mirrors is currently being tested. The research is carried out in order to clarify which material is optimal in terms of thermal expansion coefficient, manipulation, weight, and capability to be polished using the polishing techniques available in our laboratory. We are testing the following materials:
\begin{enumerate}
\item \textbf{Marble:} After intensive tests, we concluded it is not a suitable material for our specifications, especially with respect to the surface precision we aim to reach. For other applications, with more relaxed surface constraints, it remains a suitable option.
\item \textbf{Ceramics:} We tested this material with the aim of developing tools for the abrasive process (as discussed in Section\,\ref{sec:glass_mirror}
. These tools are formed by a plaster structure with hexagonal ceramic mosaics on their surface. The ceramic part is the one that comes in contact with the base material to be shaped as a mirror. We are currently working on improving the spatial layout, size, and separation of the mosaics, to maximize the tool performance.
\item \textbf{Glass:} In this project, we are mostly working with high hardness glass, such as Pyrex glass (and equivalent working recipes currently being tested at a Chilean local company), Silica (SiO2), and fused quartz, among others. Their most notable advantages are their low thermal expansion coefficient and their hardness, which make them best suited for the abrasive stage. 
A silica mandrel (high purity silicon oxide) is currently being polished. This mandrel is circular, convex, with a diameter of 15 cm and a thickness of 8 mm (Fig. \ref{fig:silica_mirror}). Even though the mandrel is convex and the resulting mirror, is obviously concave, the polishing processes of both surfaces have negligible differences. When grinding with the abrasive tool, to curve a concave shape the tool simply has to have a convex shape (and vice versa). Although this is on-going work, preliminary results suggest that given its hardness, the process takes up to four times more time when using silica (for a final product of equal dimensions).
\end{enumerate}

 \begin{figure}[H]
\begin{subfigure}{.5\textwidth}
  \centering
  \includegraphics[width=.9\linewidth]{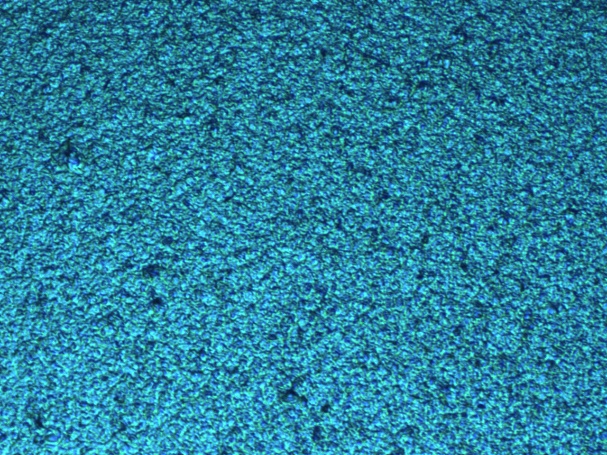}
\end{subfigure}%
\begin{subfigure}{.5\textwidth}
  \centering
  \includegraphics[width=.9\linewidth]{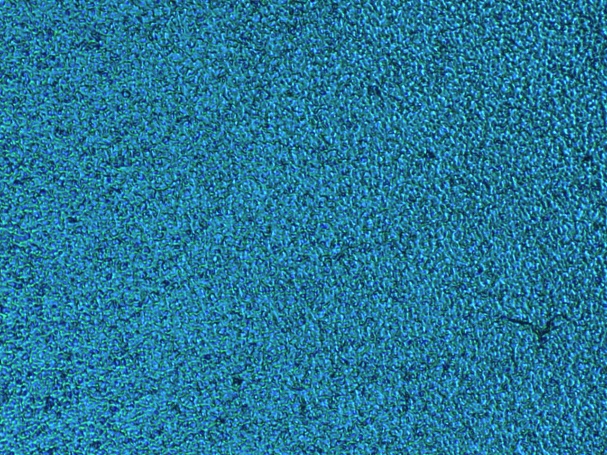}
\end{subfigure}
\caption{Microscope images of the surface of the Silica mirror after polishing process with aluminum oxide of $20$\,$\mu$m (left) and $5$\,$\mu$m (right). The area spatial scale of each image is identical, with a width of $90$\,$\mu$m.}
\label{fig:polished_surface}
\end{figure}   

\subsubsection{Polishing technique}

Given the large number of variables in the abrasive process, materials, etc., the steps on the polishing process itself have not been greatly explored 
. 
Nonetheless, from the study of the different types of tools, we found out that the hard body 
tools, with abrasive cloth attached to their surfaces, have a higher material removal rate compared to optical pitch tools. For the very same reason, this leads to a rougher surface for the end product. To correct for this issue, the tool was changed to one with an optical pitch as a contact surface and we also switched to an abrasive with lower hardness, such as powdered alumina or cerium oxide.
The use of digital roughness-meter, analogous spherometer, Ronchi test and Foucault test, allow us to test a variety of control methods.
Therefore, nowadays, we are able to combine visual inspection and a numerical measurements of the surface quality, which allows us to gain intuitive experience on what is a good or bad final surface. 

\subsubsection{Polishing abrasives}
During the abrasive process of the silica mandrel, in the polishing stage, we carried out a series of tests with Carborundum of $40$\,$\mu$m and Alumina of $50$\,$\mu$m, with the aim of measuring the final surface roughness generated using these abrasives. Before performing the test, we expected that the carborundum would generate a surface roughness greater than the alumina, due not only to the greater hardness of the former ($9.5$ Mohs compared to $9$) but also given that alumina is crushed during the abrasive process, which is not the case of carborundum. After the test, we concluded that the grain size dominates over the other characteristics of the abrasive used. We obtained a lower surface roughness with $40$\,$\mu$m carborundum. Currently, we are using tap water as a lubricant (sufficient for preliminary tests), but it will be necessary to use distilled and filtered water in the near future, to eliminate suspended particles of sizes greater than $1$\,$\mu$m that can generate surface damage during the stages of greater precision.

\subsubsection{Surface measurements	}
The Foucault (or “knife Edge”) and Ronchi tests are the standard ones used by amateur and professional telescope manufacturers to assess the optical quality of their surfaces, namely
their shape, accuracy, and surface roughness. Those two tests provide qualitative assessments of the quality of the final product, but interferometric and wavefront sensing measurements can also be performed for more quantitative results.
At those stages of development, we are currently using the Ronchi and Foucault tests, to have a first order estimate of the surface quality, and correct for any imperfections left after the polishing process. In the near future, we will combine those tests with measurements done with the Optino device (Fig.\,\ref{fig:optino}) to measure imperfections at a much finer resolution.


One should note however that the Foucault and Ronchi tests are designed for concave surfaces. Therefore, to measure the surface quality of the convex mandrels, we are currently using contact measuring elements, such as the roughmeter and the spherometer. 

\begin{figure}[H]
	\centering
	\includegraphics[width=0.45\textwidth]{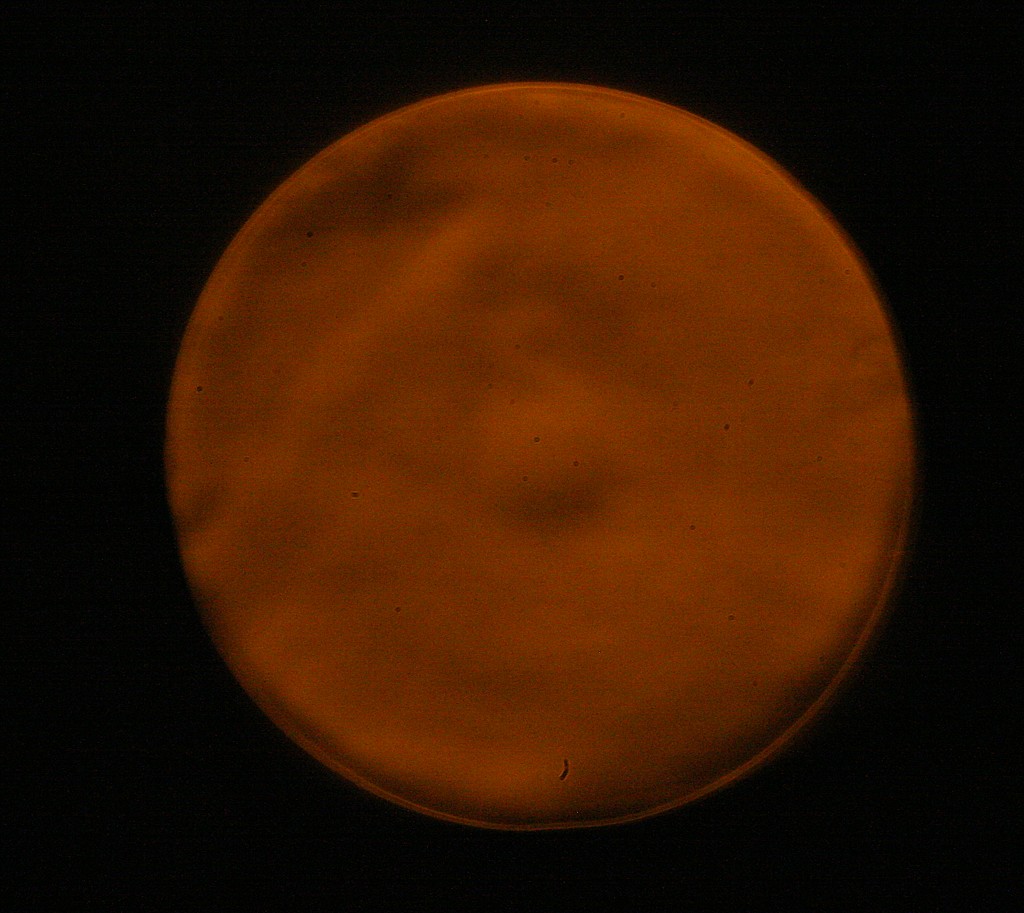}
	\includegraphics[width=0.405\textwidth]{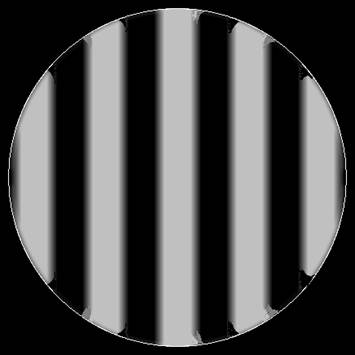}
	\caption{\textit{Left}: Foucault test showing a spheric surface with high roughness. \textit{Right}: Ronchi test showing a almost perfect spheric shape with the edges slightly elevated.}
	\label{fig:foucault_test}
   \end{figure}
   
   
\section{Summary and Conclusions}

Chile hosts the most astronomical facilities due to its pristine night sky, but is currently not a major player in the research and development for future telescope projects. Most of the mirrors are produced elsewhere than South America, which comes at a great cost and risk during transportation. Within the NPF project, we aim at remedying to this issue and making Chile a significant contributor to future missions and facilities, shifting from a ``user'' standpoint, to a ``producer'' position. This represents an important step for technological development in Chile.

The initiative to manufacture astronomical mirrors is still in its youth, the NPF project having started in September 2017. Nonetheless, in a timescale of a few months, our team has made significant progress, setting up a laboratory with existing and new equipments, testing different materials, developing a work-flow, and designing technological procedures through our experimentations. The carbon fiber mirror technology is a very promising avenue to manufacture lightweight alternative to classical glass-based mirrors. The most iconic example of similar investigation is the one performed at the ``Composite Mirror Applications'' company for the ULTRA telescope project\cite{Romeo2006}. In parallel to this innovative research, we are also developing the know-how to manufacture more classical glass-based mirrors, and we are buying additional equipments to complement the existing ones in our laboratory for those purposes. Despite the youth of the project, we already have promising and very encouraging results for both the carbon fiber and glass-based technological developments, and our technical team has shown great potential and progress in the course of a few months.




\acknowledgments 
S.\,Z.-F., A.\,B., J.\,O., L.\,P., C.\,L., N.\,S., M.\,S., P.\,E., C.\,R., H.\,H., J.\,C., and C.\,R. acknowledge financial support from the ICM (Iniciativa Cient\'ifica Milenio) via the N\'ucleo Milenio de Formaci\'on Planetaria grant. 

\bibliography{report} 
\bibliographystyle{spiebib} 

\end{document}